\documentclass[aps,prl,reprint,floatfix,superscriptaddress,amssymb,amsmath]{revtex4-2}

\usepackage{graphicx}
\usepackage{bm}
\usepackage{bbm}
\usepackage{dsfont}
\usepackage{dcolumn}
\usepackage[mathlines]{lineno}
\usepackage{color}
\usepackage{hyperref}
\usepackage{verbatim}

\begin{document}

\title{Optical Neural Networks from Coherent Transient Dynamics in Waveguide QED}

\author{Jiande Cao}
\affiliation{School of Physics, Sun Yat-sen University, Guangzhou 510275, China.}
\author{Yexiong Zeng}
\affiliation{Key Laboratory of Low-Dimensional Quantum Structures and Quantum Control of Ministry of Education, Hunan Normal University,
Changsha 410081, People’s Republic of China.}
\affiliation{Center for Quantum Computing, RIKEN, Wako-shi, Saitama 351-0198, Japan.}
\author{Franco Nori}
\affiliation{Center for Quantum Computing, RIKEN, Wako-shi, Saitama 351-0198, Japan.}
\affiliation{Department of Physics, University of Michigan, Ann Arbor, Michigan 48109-1040, USA.}
\author{Ze-Liang Xiang}
\email{xiangzliang@mail.sysu.edu.cn}
\affiliation{School of Physics, Sun Yat-sen University, Guangzhou 510275, China.}
\affiliation{State Key Laboratory of Optoelectronic Materials and Technologies, Sun Yat-sen University, Guangzhou 510275, China.}

\date{\today}

\begin{abstract}
Optical neural networks promise ultrafast, low-energy information processing by performing computation directly with photons. Current implementations, however, are largely restricted to steady-state operation and rely on high-latency electro-optical conversion for nonlinear activation. To address these limitations, we propose an all-optical fully connected neural network architecture in which the basic neuronal functions are realized by coherent transient quantum dynamics. Within this framework, phase-tunable nonlocal interference in a giant cavity implements programmable synaptic weights; an integrator operating in the bad cavity regime performs temporal summation by coherently combining sequential wavepackets; and transient Rabi dynamics of a driven two-level system provide nonlinear activation. Full-physics simulations demonstrate high classification accuracy on MNIST and colored-object recognition tasks. These results eliminate the optoelectronic activation bottleneck, reduce latency, and establish transient light-matter dynamics as a native physical resource for high-dimensional nonlinear information processing, paving the way toward fully optical neuromorphic computing.
\end{abstract}

\maketitle

\emph{Introduction.---} Artificial intelligence (AI) is driving computational demand toward the limits of the von Neumann architecture, motivating hardware paradigms that process information directly through physical dynamics rather than through sequential electronic operations~\cite{WOS:000355286600030,WOS:001468180100008,Kudithipudi2025,WOS:001565005100031,WOS:000749546400022,10.1145/3282307,rumelhart1986learning,Biamonte2017,Wang2017,Wang2018,Hoffmann2022,https://doi.org/10.1002/adma.202306818,Schuller2022}. In particular, optical neural networks (ONNs) have emerged as a promising platform to realize such physical dynamics-based computing because photons offer enormous bandwidth, low latency, minimal transmission loss, and robust coherence~\cite{WOS:001317064900006,WOS:001028306300001,Zhang2025,WOS:000809136500001,Shastri2021,WOS:001080686500002,WOS:000613520200004,Xue2024,Bandyopadhyay2024,PhysRevX.9.021032}. Most existing photonic computing platforms, however, including integrated Mach-Zehnder-interferometer networks~\cite{Shen2017,WOS:000963349400001,PhysRevApplied.21.014028} and free-space diffractive processors~\cite{ WOS:000443892700043,10.1063/5.0215752,WOS:001317064900006}, essentially operate on a steady-state regime. In these architectures, light mainly serves as a carrier of optical fields, and matrix operations are emulated through spatial linear superposition, while the transient dynamics and quantum coherence inherent in light-matter interactions remain largely unexplored~\cite{WOS:000451856000022,RevModPhys.82.1041,WOS:000307010700031,Bliokh2015,doi:10.1126/science.aaa9519,Bliokh2012,PhysRevA.95.063849}. More importantly, the lack of efficient native optical nonlinearities has led high-performance ONNs to rely on electro-optical conversions for activation, thereby compromising the all-optical advantage and introducing substantial latency and energy costs~\cite{8769881,WOS:001317064900006,WOS:000750832600001,WOS:000794035400001}.

Waveguide quantum electrodynamics (WQED) provides an natural platform for addressing these limitations~\cite{WOS:000636997400007,WOS:000401009300001,WOS:000396116600040,chernykh2025quantumopticalneuralnetworks,WOS:000476555600001}. In contrast to conventional cavity QED, which is limited by narrowband discrete modes, WQED couples quantum emitters to a one-dimensional continuum of photonic modes~\cite{PhysRevLett.95.213001,WOS:000418464600001,WOS:000655978700001,WOS:000286739100001,Kockum2019}. This setting supports broadband pulse propagation together with appreciable light-matter interaction~\cite{PhysRevLett.95.213001,WOS:000293705600006,WOS:000277945900017,WOS:000396116600040,WOS:000347229300009, WOS:000342446900051,WOS:000354347800001,WOS:000429227100001,WOS:000372399500006,WOS:000286739100001,Wallraff2004}. An even richer resource arises in the giant-atom regime, where artificial atoms are coupled to a waveguide at multiple spatially separated points rather than behaving as pointlike emitters~\cite{WOS:000554831500030,WOS:000429227100001,WOS:000494944200016,WOS:000494944200016,WOS:000612213700005,hu2024engineeringenvironmentsuperconductingqubit,PhysRevResearch.2.043184,WOS:000339973000013}. The resulting propagation and coupling phases generate nonlocal interference~\cite{WOS:000843129600001,WOS:000761167300006,WOS:000612213700005}, enabling tunable effective light-matter coupling and even decoherence-free subspaces~\cite{WOS:000554831500030,WOS:000429227100001}. Such flexible control of dissipation and transient dynamics provides a distinct route to photonic neural computation, suggesting that synaptic weighting, temporal integration, and nonlinear activation may be implemented directly via quantum dynamics rather than external electronic control~\cite{WOS:000487275900007,chernykh2025quantumopticalneuralnetworks,WOS:000476555600001}. This naturally raises a central question: can transient quantum dynamics themselves provide the \textit{linear and nonlinear} primitives required for a fully optical neural network? 

In this work, we propose a fully physical architecture for optical fully connected neural networks based on transient photon--atom dynamics. Information is encoded in the complex-valued area of coherent optical wavepackets, so that both the temporal envelope and the optical phase participate in the computation, thereby mapping classical neuron operations onto scattering and evolution in a quantum system. Three quantum mechanisms realize the required computational primitives: a giant cavity, where phase-tunable nonlocal interference enables fast control of the transmitted complex area without relying on thermo-optic modulation; a critical-gain temporal integrator, where pump-compensated bad-cavity dynamics enables near-lossless coherent accumulation of sequential pulses; and a two-level system (TLS), where transient Rabi dynamics provides an ultrafast, coherent, and intrinsically nonlinear activation mechanism acting directly on propagating wavepackets. Numerical simulations achieve classification accuracies of $97.60\%$ on MNIST and $92.32\%$ on a colored-object dataset. Furthermore, numerical results indicate strong approximation capability for both linear and nonlinear functions, as expected from universal approximation. More broadly, our results establish transient quantum dynamics in WQED as a native physical resource for the \textit{linear and nonlinear} operations underlying neural computation, thereby opening a route to fully optical neuromorphic processing beyond the steady-state paradigm.

\begin{figure}
    \centering
    \includegraphics[width=1\linewidth]{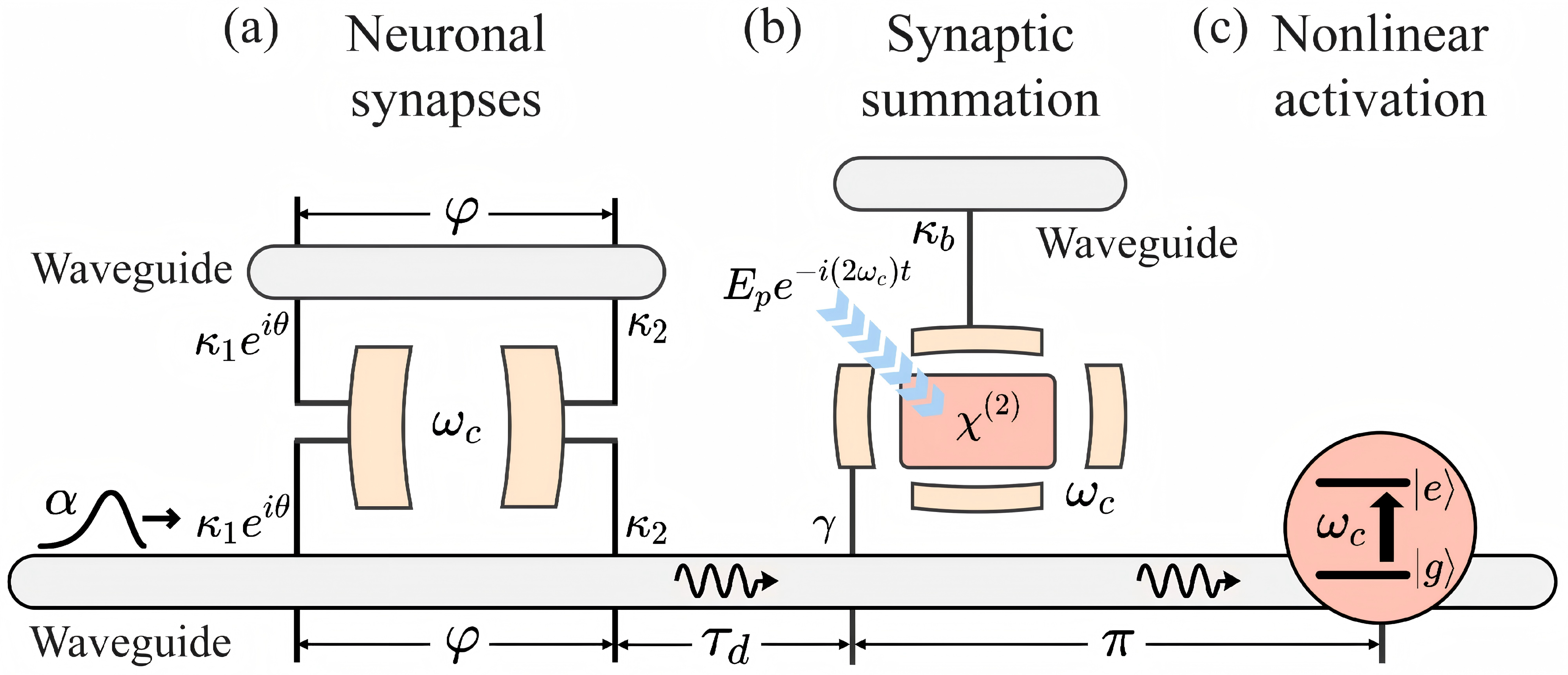}
    \caption{Physical implementation of a neuron in the WQED architecture. 
    (a) Synaptic-weighting module based on giant-cavity interference. The lower and upper waveguides carry the input signal  $c_{\rm sig}$ and dissipated energy $c_{\rm diss}$, respectively, with identical coupling configurations $\kappa_1e^{i\theta}$ and $\kappa_2$.
    (b) Pump-assisted temporal integrator. The main and auxiliary cavities, supporting modes $a$ and $b$, operate at frequency $\omega_c$ with decay rates $\gamma$ and $\kappa_b$, respectively. A classical pump at $\omega_p=2\omega_c$ drives the $\chi^{(2)}$ nonlinear medium and enables coherent accumulation of sequential pulses.
    (c) Nonlinear activation realized via a TLS with transition frequency $\omega_c$.}
    \label{neuron}
\end{figure}

\emph{Model.---}
We consider a WQED architecture composed of three modules: a giant-cavity weighting element, a lossless temporal integrator, and a TLS--connected by propagating wavepackets in a chiral waveguide, as shown in Fig.~\ref{neuron}. Specifically, phase-controlled nonlocal interference in the giant cavity sets the synaptic weighting of each wavepacket, a critical-gain integrator in the bad-cavity regime coherently accumulates sequential pulses through pump-compensated dissipation, and the transient Rabi response of the TLS provides an intrinsically nonlinear transformation of the propagating wavepacket.

To describe all three modules using a unified dynamical framework, we encode each input value in the complex area of a wavepacket,
\begin{equation}
\mathcal{A} \equiv \int \alpha(t) dt,
\label{area}
\end{equation}
where $\alpha(t)$ is the temporal envelope of the wavepacket in the interaction picture of the system. In the frequency domain, $\mathcal{A}$ is the zero-frequency component of the envelope. The input data to be processed. i.g., image pixel values, are mapped onto exponential decaying wavepackets $\alpha(t)$ and injected into the chiral waveguide.

We first implement synaptic weighting via nonlocal interference in a giant cavity [see Fig.~\ref{neuron}(a)]. The giant cavity is coupled to two chiral waveguides, each at two spatially separated points. By tuning the local phase $\theta$ at the first coupling point relative to the propagation phase $\varphi$, the transmission amplitude becomes a tunable synaptic weight. Temporal summation is then implemented by a cavity integrator [see Fig.~\ref{neuron}(b)], where the two cavities interact through a $\chi^{(2)}$ nonlinear medium driven by a pump of amplitude $E_p$. Finally, nonlinear activation is provided by the transient Rabi response of a strongly driven TLS [see Fig.~\ref{neuron}(c)].

All calculations are performed in the interaction picture. We set all cavity and atomic transition frequencies to be resonant to $\omega_c$, which provides a unified interaction-picture description of the three modules and a consistent treatment of wavepacket propagation. For the weighting and summation stages, the interaction Hamiltonian reads 
\begin{equation}
\begin{aligned}
    H_{I}(t)&=  i \int d \omega \bigg\lbrace\big[\frac{\Lambda}{\sqrt{2 \pi}} c_{\text {sig }}^{\dagger}(\omega) a_{c}+\frac{\Lambda}{\sqrt{2 \pi}} c_{\text {diss }}^{\dagger}(\omega) a_{c} \\
    & +\sqrt{\frac{\gamma}{2 \pi}} e^{i \omega \tau_{d}} c_{\text {sig }}^{\dagger}(\omega) a +\sqrt{\frac{\kappa_{b}}{2 \pi}} d^{\dagger}(\omega) b\big]e^{i\left(\omega-\omega_{c}\right) t} \\
    & +E_p \chi^{(2)}a^{\dagger} b^{\dagger}-\text{h.c.}\bigg\rbrace,
\end{aligned}
\label{H}
\end{equation}
with $\Lambda \equiv \sqrt{\kappa_2} + e^{i\varphi}\sqrt{\kappa_1}\,e^{i\theta}$, $\hbar=1$, and $\tau_{d}=2\pi$ . After summation, the pump is switched off, the stored wavepacket is released from the main cavity, and the output pulse is sent to the TLS for activation.

\emph{Neuronal synapses.---}Synaptic weighting is realized through nonlocal interference in a giant cavity coupled to two chiral waveguides, thereby enabling precise control of the complex-valued area of the output wavepacket in the signal channel. Using the SLH formalism~\cite{WOS:000429227100001} and input-output theory~\cite{PhysRevA.31.3761}, 
we find that the giant-cavity weighting module is characterized by the complex-valued area transmittance (see Suppl. Mat., SM),
\begin{equation}
T \equiv \mathcal{A}_{\rm out}/\mathcal{A}_{\rm in} = |T|e^{i\psi},
\end{equation}
where $\mathcal{A}_{\rm in}$ and $\mathcal{A}_{\rm out}$ denote the complex-value areas of the input and output  wavepackets, respectively. We take the synaptic weight as $\vert T\vert$, controlled solely by the coupling phase $\theta$. The accompanying phase shift $\psi(\theta)$ can be compensated at the input of the next summation module, thereby reducing the operation to multiplication by a real scalar.

Importantly, within the complex-area encoding scheme, the weight-calibration curve depends only weakly on the detailed wavepacket shape. As a result, waveform distortion accumulated over multiple layers has little influence on the implemented weight (SM), enabling fully optical operation without optoelectronic conversion.

\emph{Synaptic summation.---}Temporal summation is realized via coherent accumulation of a sequence of weighted pulses. We thus operate the cavity in the bad-cavity regime and introduce a gain $G$ to compensate the intrinsic loss $\gamma$, i.e., $G \approx \gamma$. At this critical point, the cavity acts as a temporal integrator that stores the incoming pulse sequence as a sum of the complex-valued areas of the wavepackets [see Fig.~\ref{neuron}(b)].

The intracavity field obeys the quantum Langevin equations (SM)
\begin{align}
    \dot{a}(t) &\approx - \sqrt{\gamma}c_{\rm in}(t) -\sqrt{G}\,d_{\mathrm{in}}^\dagger(t) ,
\label{sum1}
\end{align}
where $c_{\rm in}(t)$ is the weighted output from the giant-cavity module; $G\equiv4|\Omega|^2/\kappa_b$ denotes the effective gain, with $\Omega=\chi^{(2)}E_p$ the effective pump strength, and  $-\sqrt{G}\,d_{\mathrm{in}}^\dagger(t)$ is the effective noise. At critical gain, the cavity thus integrates the input signal up to an additive noise term. Such noise can play a constructive role by acting as stochastic fluctuations that help the system escape poor local minima, providing a form of physical regularization.

After the pulse sequence has been integrated, the stored field is released as $\beta(t) = -\sqrt{\kappa_{r}} \alpha_{\rm sum} e^{-\kappa_{r}t/{2}}$, where $\kappa_{r}$ is the coupling rate to the output waveguide and  $\alpha_{\rm sum}$ is the total stored complex-valued area, including the noise contribution. After an additional phase shift $\pi$, the released pulse is sent to the TLS for nonlinear activation.

\emph{Nonlinear activation.---}Activation is provided by a TLS coupled to a waveguide [see Fig.~\ref{neuron}(c)]. The input wavepacket $\epsilon_{\rm in}(t)$ is the pulse $\beta(t)$ released from the integrator after a phase shift of $\pi$. As it traverses the TLS, dipole coupling induces atomic radiation that interferes with the transmitted field, producing the output pulse $\epsilon_{\rm out}(t)$. Because the TLS is saturable and undergoes transient Rabi dynamics under strong driving, this input-output map is intrinsically nonlinear.

\begin{figure}
    \centering
    \includegraphics[width=1\linewidth]{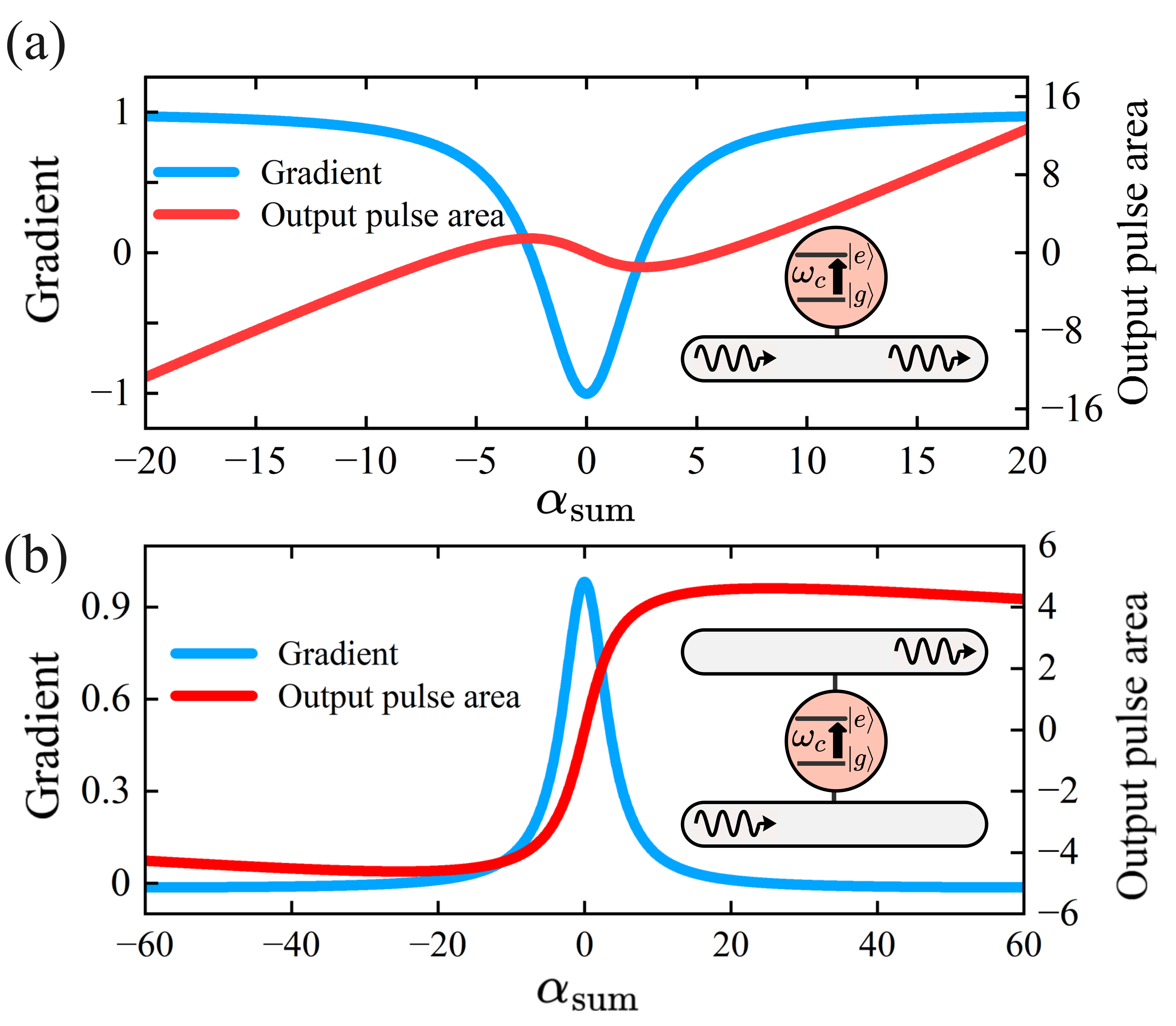}
    \caption{Nonlinear activation function. 
    (a) Activation based on a TLS, where the incident wavepacket traverses the TLS. Red: complex area of the output wavepacket versus the complex area of $\alpha_{\rm sum}$. Blue: gradient of the activation function. 
    (b) Activation based on a multiple waveguides structure, where an incident wavepacket from one waveguide produces an outgoing wavepacket in the other waveguide. Red: complex area of the output wavepacket versus the complex area of $\alpha_{\rm sum}$. Blue: corresponding gradient of the nonlinear activation function.}
    \label{nonlinear activation}
\end{figure}

The dynamics of the TLS are governed by the Heisenberg-Langevin equations
\begin{align}
\!&\frac{d}{dt}\sigma_-(t) = -\frac{\Gamma_{\rm at}}{2}\sigma_-(t) + \sqrt{\Gamma_{\rm at}}\,\epsilon _{\rm in}(t)\,\sigma_z(t) , \\
\!&\frac{d}{dt}\sigma_z(t) = -\Gamma_{\rm at}\left[1+\sigma_z(t)\right] - 2\sqrt{\Gamma_{\rm at}}\,\mathrm{Re}\!\left[\epsilon_{\rm in}^*(t)\,\sigma_-(t)\right]
\end{align}
where operators $\sigma_-(t)$ and $\sigma_z(t)$ describe the dipole amplitude and the population inversion of the TLS, respectively. $\Gamma_{\rm at}$ is the TLS-waveguide coupling strength. The nonlinearity originates from the state-dependent driving term
$\epsilon_{\rm in}(t)\sigma_z(t)$ in the equation for $\sigma_-(t)$. Then, the output field is $\epsilon_{\rm out}(t) = \epsilon_{\rm in}(t) + \sqrt{\Gamma_{\rm at}}\,\sigma_-(t)$. The complex area of $\epsilon_{\rm out}(t)$ thus defines the activation function acting on the integrated input. As shown in Fig.~\ref{nonlinear activation}, it suppresses small inputs while leaving large inputs nearly unchanged. 

\begin{figure*}
    \centering
    \includegraphics[width=1\linewidth]{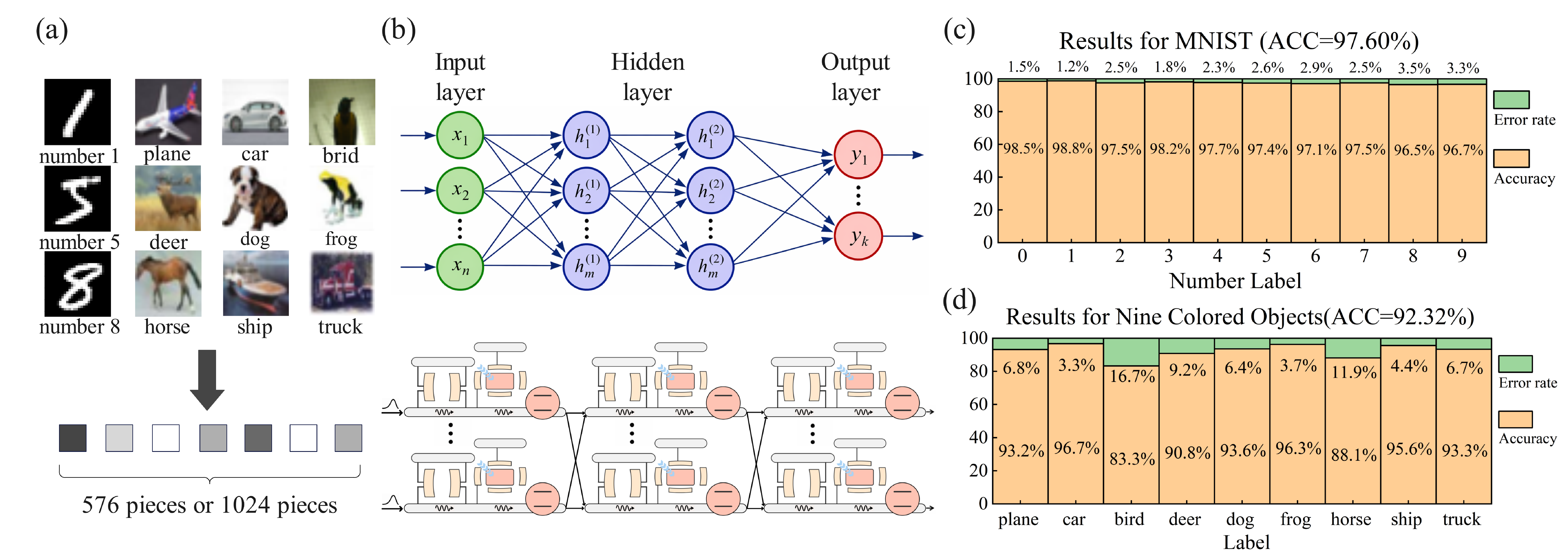}
    \caption{Architecture and classification performance of the quantum-optical neural network. 
    (a) Representative training samples for the two classification tasks: MNIST and colored-object recognition. The images are flattened into 576 and 1024-dimensional vectors, respectively. Each vector component is encoded as an exponentially decaying wavepacket and injected into the network. 
    (b) Quantum optical neural network architecture. The injected wavepackets form the input layer, intermediate neuron layers form the hidden layers, and the output layer encodes the predicted class probabilities. Both tasks use two hidden layers. The network sizes are $576$--$128$--$128$--$10$ for MNIST and $1024$--$256$--$256$--$9$ for nine-object classification. 
    (c) Classification results for the MNIST task, yielding an accuracy of $97.60\%$. (d) Classification results for the nine-colored-object task, yielding an accuracy of $92.32\%$.}
\label{ARCHITECTURE AND RESULT}
\end{figure*}

An alternative implementation is realized in a multiwaveguide structure, where the atom couples to separate signal and output waveguides. A wavepacket injected through the signal waveguide generates an output pulse in the output waveguide, which can be routed directly to the next neuron. The resulting nonlinear function exhibits a \textit{tanh-like} activation profile, offering a distinct physical realization of the activation stage.

\emph{Architecture and results.---}By interconnecting the physical neurons described above through a waveguide network, we construct a quantum-optical neural network (QONN). In the forward propagation, classically preprocessed image data are encoded into the complex-valued areas of exponentially decaying wavepackets $\mathcal{A}_{\rm in}$, normalized to the operating range of the TLS, and injected into the waveguide at regular time intervals. As the pulse propagates through the weighting and summation modules, the tunable transmittance $T(\theta)$ implements the linear operation $z=\sum_i w_i x_i$. The resulting pulses then drives the TLS, whose transient nonlinear response realizes the nonlinear activation $a=\sigma(z)$.

After propagation through multiple layers, the distribution of output wavepacket complex-valued areas across the final channels defines the network's predicted class probabilities. For the $i$ th output neuron, we read 
\begin{equation}
y_i=\int \left \langle \epsilon_{\rm out,i}(t) \right \rangle  dt,
\end{equation}
and normalize the result to obtain $y_{\mathrm{pred},i}$, which define the predicted classification probabilities. Training is performed by backpropagation~\cite {rumelhart1986learning}: the cross-entropy loss is computed from the target labels $y_{\mathrm{true},i}$ and the predicted probability $y_{\mathrm{pred},i}$, and gradients are obtained via the chain rule to update the coupling phase $\theta$.

We adopt an \emph{in-situ} training strategy~\cite{WOS:000439429000016}, in which forward propagation is carried out on the physical network, while parameter updates are performed on a classical computer. We benchmark the architecture on two tasks: handwritten-digit recognition and nine-colored-object classification [see Fig.~\ref{ARCHITECTURE AND RESULT}(a)]. Following classical convolution and pooling, the processed image data are encoded into exponentially decaying wavepackets and injected into the QONN. 

After training, the network reaches classification accuracies of 97.60\% on MNIST and 92.32\% on nine-colored-object classification [see Figs.~\ref{ARCHITECTURE AND RESULT}(c, d)]. These results show that giant-cavity interference, temporal integration, and nonlinear TLS dynamics can be combined into a multilayer computing architecture operating directly on propagating wavepackets, and establish coherent transient photon dynamics as a viable physical resource for neuromorphic computation. 

With the multiwaveguide implementation of the activation stage, we obtain 97.51\% accuracy on MNIST. This structure preserves a strong nonlinear response for large input signals, and exhibits a \textit{tanh-like} activation profile. 

In practical physical devices, some components may exhibit errors. So we further test the neurons' robustness to parameter drift and phase noise, modeling fabrication imperfections and thermal fluctuations. Even after only 50 training epochs, the network remains highly accurate under strong perturbations: with a 25\% deviation of the gain $G$ from the critical point, the accuracy stays near 95\%, and with independent Gaussian phase noise of zero mean and 32° variance on each wavepacket, it still reaches 94.33\% (see Suppl Mat, SM). These results support the physical feasibility of the architecture.

\emph{Feasibility and Architectural Features.---}Our architecture is compatible with the state-of-the-art superconducting quantum platforms, where its three neuronal primitives can be mapped onto circuit-QED components. Giant-cavity interference for synaptic weighting can be implemented with a superconducting resonator coupled to a microwave transmission line at multiple points, with phase tunability provided by a SQUID~\cite{WOS:000256196600092}. The lossless temporal integrator can be realized through a driven $\chi^{(2)}$ nonlinearity in the bad-cavity regime, as in a Josephson parametric converter operated near critical gain~\cite{Bergeal2010,Bergeal2010Nature}. Nonlinear activation arises from transient Rabi dynamics in a transmon qubit coupled to a 1D waveguide. For megahertz-scale couplings, the relevant dynamics unfold on nanosecond time scales, indicating that the proposed scheme is experimentally accessible with existing superconducting technology.

More broadly, the architecture suggests a distinct route to optical computation in which information processing is carried out directly through coherent transient dynamics, rather than through sequential electronic control~\cite{10.1145/3282307,WOS:000444077900009}.

A central design element is the encoding of high-dimensional inputs as a time-ordered sequence of exponentially decaying wavepackets injected through a single waveguide. Compared with parallel spatial encoding, this time-division strategy reduces footprint and hardware complexity while retaining efficient linear summation. Signed values are encoded in the complex-valued area of each wavepacket, with a $\pi$ phase shift representing negative amplitudes, so that signed linear operations can be implemented directly without auxiliary differential or bias circuits. Residual intracavity energy is removed via phase-controlled decoupling into a dissipation channel, thereby suppressing intersymbol interference in high-$Q$ resonators, while nonlinear activation is implemented via the TLS. By eliminating optoelectronic conversion at the activation stage, this proposal enables compact, signed, and fully optical information processing. 

\emph{Conclusion.---}We proposed and demonstrated a neuromorphic architecture based on transient quantum optical dynamics. Unlike traditional optical computing approaches that rely on the linear superposition of steady-state intensities, our proposal encodes information in the complex-valued areas of coherent wavepackets. By exploiting giant-cavity interference and bad cavity dynamics, we physically implement synaptic weighting and temporal integration. Notably, the system is not merely limited by noise but benefits from it: the noise plays a constructive role analogous to the stochasticity in stochastic gradient descent algorithms, facilitating the escape from local minima during training. Furthermore, we leverage the quantum optical response of a coupled TLS to directly realize nonlinear mapping in the optical domain, achieving high classification accuracies on both tasks.

Our architecture is compatible with current integrated photonics platforms, where giant-cavity couplings and fast modulation are experimentally accessible. This work opens new possibilities for future ultrafast and energy-efficient neuromorphic processors harnessing the transient physics of light-matter interactions.

\section*{Acknowledgement}

This work is supported by the National Natural Science Foundation of China (Grant No. 12375025). F.N. is supported in part by: the Japan Science and Technology Agency (JST) [via the CREST Quantum Frontiers program Grant No. JPMJCR24I2, the Quantum Leap Flagship Program (Q-LEAP), and the Moonshot R\&D Grant Number JPMJMS256E].


\bibliography{reference_onnwqed}


\end{document}